\documentstyle[aps,epsf,psfig]{revtex} 
\newcommand{\be}[1]{\begin{equation}\label{#1}}
\newcommand{\ee}{\end{equation}}

\renewcommand{\vec}[1]{{\bf #1}}
\newcommand{\vecgr}[1]{\mbox{\boldmath{$ #1$}}}
\newcommand{\Tr}{\mbox{Tr}}

\newcommand{\N}{\mbox{I}\!\mbox{N}}
\newcommand{\R}{\mbox{I}\!\mbox{R}}
\newcommand{\K}{\mbox{I}\!\mbox{K}}
\newcommand{\tK}{\tilde{\mbox{I}\!\mbox{K}}}

\newcommand{\bino}[2]{\left(\begin{array}{c} #1\\#2\end{array}\right)}
\renewcommand{\v}{\vec{v}}

\begin{document}
\title{
Spectral statistics for unitary transfer matrices of binary graphs 
}
\author{ Gregor Tanner} 
\address{
School of Mathematical Sciences
\footnote{e-mail: gregor.tanner@nottingham.ac.uk}\\
Division of Theoretical Mechanics\\
University of Nottingham\\
University Park, Nottingham NG7 2RD, UK
}

\maketitle

\begin{abstract}
Quantum graphs have recently been introduced as model systems to study the 
spectral statistics of linear wave problems with chaotic classical limits.
It is proposed here to generalise this approach by considering arbitrary, 
directed graphs with unitary transfer matrices.
An exponentially increasing contribution to the form factor is identified
when performing a diagonal summation over periodic orbit degeneracy classes. 
A special class of graphs, so--called binary graphs, is studied in more detail.
For these, the conditions for periodic orbit pairs to be correlated  
(including correlations due to the unitarity of the transfer matrix) can be 
given explicitly.
Using combinatorial techniques it is possible to perform the summation
over correlated periodic orbit pair contributions
to the form factor for some low--dimensional cases.
Gradual convergence towards random matrix results is observed when increasing
the number of vertices of the binary graphs.\\

\noindent
{\normalsize Submitted to {\em Journal of Physics A};\\
Version: 30th November 1999.}
\end{abstract}
\section{Introduction}
\label{sec:sec1}
Universality in spectral statistics has be established numerically and
experimentally for a wide range of linear wave problems ranging from quantum 
systems (Bohigas et al (1984)) to acoustic (Ellegaard et al (1996)) and 
microwave cavities (Alt et al (1997, 1999)) in two and three dimensions as 
well as quantum maps (Saraceno and Voros (1994)) and quantum graphs (Kottos 
and Smilansky (1997, 1998)), see also Guhr et al (1998) for a recent review. 
The universality classes are accurately described by random matrix theory (RMT) 
even though ensemble average is not performed when considering 
spectra of individual wave problems. This fundamental puzzle is not understood
until today and indicates that the RMT--limit is reached under more
general conditions than assumed by Wigner, Mehta, Dyson and others 
(see e.g.\ Mehta (1991)) in the original derivation of RMT -- results. 

A few basic facts are well established by now: wave systems, whose spectral 
statistics follow the RMT--result for Gaussian unitary or orthogonal
ensemble (GUE or GOE) have in common that 
\begin{itemize}
\item[a)] time propagation (discrete or continuous) is a linear, unitary 
transformation;
\item [b)] the dynamics of the underlying classical system is chaotic; this 
implies in particular that the classical Perron-Frobenius operator has an 
isolated largest eigenvalue equal to one, positive Liapunov exponent and an 
exponentially increasing number of periodic orbits;
\item [c)] there are no systematic periodic orbit length degeneracies other 
than those enforced by the symmetries of the classical dynamics and the 
unitarity of the wave propagation.
\end{itemize}
The last point is kept vague deliberately and refers to systems which 
fulfill condition (a) and (b) but are known to deviate from RMT due to number 
theoretical periodic orbit degeneracies; examples are the cat map 
(Hannay and Berry (1980), Keating (1991 a,b)) and
arithmetic billiards of constant negative curvature (Bogomolny et al (1997)). 
I will come back to this point in the next sections.

A direct consequence of (b) is the so-called Hannay--Ozorio de 
Almeida (HOdA) sum rule (Hannay and Ozorio de Almeida (1984), Berry (1985)), 
which enables one to derive universality of the spectral two point correlation 
function in the long range limit. Considerable progress in 
understanding the universality of spectral statistics for individual systems 
beyond the HOdA-sum
rule has been made only recently by studying quantum graphs.
In a series of papers Smilansky and coworkers demonstrated numerically that 
quantum graphs indeed obey RMT-statistics (Kottos and Smilansky (1997, 1999)); 
they were also able to calculated the full form factor, i.e., the Fourier 
transform of the spectral two point correlation function, in terms of periodic 
orbits 
for a specific set of graphs with $2\times 2$ unitary transfer matrices 
(Schanz and Smilansky (1999)) and reproduced Anderson localisation
from periodic orbit theory in a similar model (Schanz and Smilansky (1999a)). 
Deviations from universal statistical behaviour for a special set of graphs -- 
so-called star-graphs -- could be explained in leading order by Kottos and 
Smilansky (1999), a systematic way to calculate higher order corrections 
has been developed by Berkolaiko and Keating (1999). 

The main advantage in studying quantum graphs is that one can construct a
wide variety of systems with exact periodic orbit trace formulae
(in contrast to, for example, semiclassical periodic orbit trace formulae,
see Gutzwiller (1990)). Discrete time propagation on a graph corresponds to 
a unitary transformation in
terms of a finite dimensional matrix and periodic orbit lengths are
build up by a finite number of rationally independent length
segments. The exactness of the trace formula circumvents problems due to,
for example, semiclassical errors present in periodic orbit trace formulae 
for general quantum systems with continuous classical limit. 
Semiclassical approximations do in general not preserve 
unitarity of the quantum propagation which leads to exponentially growing 
error terms in the long time limit (Keating (1994), Tanner (1999)). 
Periodic orbit length correlations beyond the classical HOdA-sum rules 
can furthermore be studied in graphs in detail without referring to 
approximations; such correlations are predicted to 
exist due to the presence of spectral universality (Argaman et al (1993)).

The quantisation procedure for graphs chosen by Kottos and Smilansky 
(1997, 1999) implies certain restrictions on the topological structure 
of the graph. Solving a one-dimensional Schr\"odinger equation on 
the connections (or edges) between vertices with various 
boundary conditions calls for the possibility of backscattering; the
underlying graph must therefore be undirected, i.e., the possibility 
to go from vertex $i$ to vertex $j$ implies that the reversed 
direction from $j$ to $i$ also exists. 

In the following I will broaden the picture by considering unitary 
matrices in general. I will identify a unitary matrix as a transfer
matrix (or `wave propagator') on a directed graph with exact periodic 
orbit trace formula. The corresponding
classical system is, as for quantum graphs, given by the dynamics on a 
probabilistic network. Such a construction has a priori, and again like 
for quantum graphs, no semiclassical limit in the sense that the classical
dynamics does not remain the same when increasing the matrix dimension 
(or the size
of the graph). This is, however, not a prerequisite when looking at
the conditions (a) - (c); one can indeed easily construct graphs
and corresponding unitary transfer matrices which fulfill the conditions
above. The main motivation in generalising the concept of quantum graphs
lies in the possibility to study a much wider class of graphs including 
in particular directed graphs. This freedom will be used in section 
\ref{sec:sec3}, \ref{sec:sec4} to consider a special set of graphs, 
so--called binary graphs. Unlike for quantum graphs, the unitary transfer 
matrix of a directed graph can not be written as a function of a wavenumber 
$k$ in general and does not have a quantum 
spectrum. Like for quantum maps, one studies instead the statistics of the 
spectrum of eigen-phases of the unitary matrix.\\

I will introduce some basic notations for graphs in section 
\ref{sec:sec2} and will define edge and vertex staying rates 
as well as periodic orbit degeneracy classes. An exponentially 
increasing contribution to the form factor is identified when 
performing a diagonal summation over degeneracy classes.
I will then focus on 
balanced, directed (binary) graphs with unitary transfer matrices. 
The form factor can here be written in terms of a periodic orbit length 
degeneracy function. This functions will be derived explicitly 
for binary graphs with up to 6 vertices in section \ref{sec:sec3}. 
Exponentially increasing contributions to the form factor are identified;
these contributions alternate in sign and balance each other in a delicate 
way to lead to an expression for the form factor close to the RMT - 
result. The periodic orbit form factor for graphs with up to 
32 vertices is calculated in section \ref{sec:sec4}
by counting the periodic orbit degeneracies directly. Convergence 
of the periodic orbit expressions towards the RMT -- result 
is observed for graphs with and without time reversal
symmetry; this gives rise to the hope that a periodic orbit
theory may indeed be able to resolve universality of spectral
statistics in the limit of large vertex - numbers.

\section{Graphs and unitary transfer matrices}
\label{sec:sec2}
\subsection{Introduction and notation}
A directed graph (digraph) $G$ consists of set of vertices $V(G)$ 
connected by a set of edges $E(G)$. An edge leading from a vertex 
$i$ to a vertex $j$,  ($i, j \in V(G)$), will be denoted ($ij$) and 
the ordering of the pair is important. I will mainly deal with 
directed graphs here and will omit the specification `directed' in the 
following. The order of the graph is given by the number of vertices 
$N = |V(G)|$, and $M = |E(G)|$ is the number of edges. A graph can be 
characterised by its $N\times N$ adjacency matrix $\vec{A}(G)$ 
being defined here as
\[ a_{ij} = \left\{ \begin{array}{ll} 1 &\mbox{if} \quad ij \in E(G)\\
                                          0 & \mbox{otherwise} 
                     \end{array} \right. \; ;\]
the vertices $i,j \in V(G)$ may be labeled from 0 to $N-1$
for convenience.  A real or complex $N\times N$ matrix $\vec{T}(G)$ 
will be a called a {\em transfer matrix} of $G$ if 
\[ t_{ij} = 0 \; \Leftrightarrow \; a_{ij} = 0. \]
A real transfer matrix $\vec{T}^{cl}(G)$ which preserves probability, i.e.\
\be{prob}
 \sum_{j=0}^{N-1} t^{cl}_{ij} = 1 \quad \forall i \in V(G),\, t_{ij} \in \R
\ee
is called a {\em classical transfer matrix} in what follows. 
$\vec{T}^{cl}$ is the analogue of the classical transfer or 
Frobenius--Perron
operator for dynamical systems with continuous configuration
space and describes the discrete time evolution of an $N$ -- dimensional 
vertex density vector $\vecgr{\rho}$ according to 
\[ \rho_j(n+1) = \sum_{i=0}^{N-1} t^{cl}_{ij}\; \rho_i(n), \quad 
n \in \N \; .\] 
A matrix element $t^{cl}_{ij}$ corresponds thus to the transition
probability going from vertex $i$ to $j$. The classical 
transfer matrix has a largest eigenvalue equal to one; the graph is
ergodic if there exists a walk or path from $i$ to $j$ for every 
vertex $i$ and$j$. A graph is 'chaotic' (and thus necessarily 
ergodic) if the modulus of the second largest eigenvalue is  
smaller than one. This means, an initial  density vector $\vecgr{\rho}(0)$ 
converges exponentially fast towards an equilibrium state 
$\tilde{\vecgr{\rho}}$  which is the eigenvector corresponding to the 
largest eigenvalue of $\vec{T}^{cl}$. 

A periodic orbit of period $n$ on a graph is a walk on the graph 
which repeats after $n$ steps. Each periodic orbit can be labeled in 
terms of a vertex symbol code $(v_1v_2 \ldots v_n) = \v$ given by the 
vertices $v_i \in V(G)$ visited along the walk with $v_iv_{i+1} \in E(G),
\forall i=1,n-1$ and $v_nv_1 \in E(G)$. We will denote the
set off all periodic orbits of period $n$ as ${\cal PO}_n(G)$. \\

I will in the following focus on unitary transfer matrices $\vec{T}$. 
The `classical' dynamics corresponding
to the `wave propagation' on the graph described by the unitary 
matrix $\vec{T}$ is then given by the classical transfer matrix 
$\vec{T}^{cl}$ with $t^{cl}_{ij} = |t_{ij}|^2$. The unitarity of $\vec{T}$ 
ensures the probability conservation, Eq.\ (\ref{prob}), for $\vec{T}^{cl}$
and the equilibrium state is the uniform density vector 
$\tilde{\vecgr{\rho}} = (1,1,\ldots 1)$. The complex non-zero matrix 
elements of 
$\vec{T}$ may be written as $t_{ij} = r_{ij} e^{i L_{ij}}$ and one 
identifies $L_{ij}$ with the length of an edge ($ij$) and $r_{ij}^2 
= t^{cl}_{ij}$ is the classical transition probability. 

The conditions (a) -- (c) in section \ref{sec:sec1} are fulfilled if the
graph is chaotic and the phases $L_{ij}$ are not rationally related apart 
from conditions enforced due to unitarity. The spectrum
of $\vec{T}$ and the periodic orbits in the graph are furthermore related
by an exact trace formula; the density of states for the eigenphases
$\{\theta_i\}_{i=1,N}$ of $\vec{T}$ is given as
\be{density}
d(\theta,N) = \sum_{i=1}^N \delta(\theta - \theta_i)
= \frac{N}{2\pi} + \frac{1}{\pi}{Re}\sum_{n=1}^{\infty}
\Tr \vec{T}^n e^{-i n\theta}
\ee
and the traces $\Tr \vec{T}^n$ can be written as sum over
all periodic orbits of period $n$ in the graph, i.e.\
$\Tr \vec{T}^n = \sum_{\v\in {\cal PO}_n} A_{\v} e^{i L_{\v}}$.
The amplitude $A_{\vec{v}}$ is the product over the transition
rates $r_{v_iv_{i+1}}$ along the path and $L_{\vec{v}}$ corresponds to
the total length of the periodic orbit.

The spectral measure studied in more detail in this paper is the
so--called spectral form factor $K(\tau,N)$; it is the Fourier transformed 
of the two point correlation function 
\[R_2(x,N) = \frac{4\pi^2}{N^2} \langle d(\theta) d(\theta + 2\pi x/N)
\rangle \] 
and the average $\langle .\rangle $ is taken over the $\theta$ -- interval $[0,2\pi]$. The 
form factor written in terms of periodic orbits has the form (see e.g.\
Tanner (1999))
\be{K1}
K(\tau,N) = \frac{1}{N} \langle |\Tr \vec{T}^{n}|^2 \rangle_{\Delta \tau}
= \frac{1}{N} \langle \sum_{\v,\v'\in {\cal PO}_n} 
A_{\v} A_{\v'} e^{i ( L_{\v} - L_{\v'})}\rangle_{\Delta \tau}\; 
\ee
with $\tau$ taking on the discrete values $\tau = n/N$ and further 
averaging over small intervals $\Delta \tau$ is performed.
Most periodic orbits of the graph will be uncorrelated and the
corresponding periodic orbit pair contributions will vanish after performing
the $\tau$ -- average. There are, however, 
correlations in the periodic orbit length spectrum which lead to systematic
deviations from a zero mean; the most obvious one is between orbits which 
are related by cyclic permutation of the vertex code $\v$. The sum over
those pairs of orbits leads to the HOdA -- sum rule and describes the 
linearised behaviour of 
$K(\tau)$ for $\tau \to 0$ (Berry (1985)). One can immediately identify 
another  
class of exactly degenerated orbits on graphs; this is the set of periodic
orbits which passes through each edge the same number of times but not
necessarily in the same order. After defining the so-called 
{\em edge staying rates} $q_{ij}$ of $\v$ as the number of times a give 
orbit $\v$ visits a certain edge (ij), i.e.
\be{edge_stay} 
q_{ij}(\v) = \sum_{l=1}^{n} \delta_{i,v_l}\delta_{j,v_{l+1}} 
\qquad (ij)\in E(G) \quad \v \in {\cal PO}_n\;  ,
\ee
one can write the length $L_{\v}$ and the amplitude $A_{\v}$ of an orbit 
$\v$ on a graph as
\[  L_{\v} = \sum_{ij\in E(G)} q_{ij}(\v) L_{ij}\; , \qquad
 A_{\v} = \prod_{ij\in E(G)} r_{ij}^{q_{ij}(\v)} .\]
Periodic orbits whose symbol string gives rise to the same edge staying rate
vector $\vec{q} = (\{q_{ij}\}_{ij\in E(G)})$ coincide in length $L_{\v}$ 
and amplitude $A_{\v}$; these orbits will be called 
{\em topologically degenerate}. The set of all topologically degenerated orbits 
will be called a {\em degeneracy class} (Berkolaiko and Keating (1999)).
The number of orbits in a given  degeneracy class represented by the 
$M$ dimensional edge staying rate vector $\vec{q}$ (with $M$, the number 
of edges of the graph) will be denoted the 
{\em (periodic orbit length) degeneracy function} $P_n(\vec{q};G)$, i.e.\
\be{deg-func} 
P_n(\vec{q};G) = \left|\{\v \in {\cal PO}_n | \,q_{ij}(\v) = q_{ij}, \; 
\forall ij \in E(G)\}\right|\; .
\ee
The orbits related by cyclic permutation of the symbol code are obviously 
in the same degeneracy class.

The traces of $\vec{T}$ which enter the density of states (\ref{density}) 
can thus be rewritten as
\be{tr-geo}
\Tr \vec{T}^n = \sum_{\vec{q} \in \K_n(G)} 
P_n(\vec{q}) A_{\vec{q}} e^{i L_{\vec{q}}}\, ,
\ee
and $\K_n(G) \subset \N_0^{M}$ represents the subset of the M-dimensional 
integer lattice $\N_0^{M}$ containing all the possible edge staying rate 
vectors $\vec{q}$ which correspond to periodic orbits 
of period $n$ of the graph $G$. Determining the lattice $\K_n(G)$ and thus the
possible degeneracy classes as well as the degeneracy function is the
main problem when studying periodic orbit length correlations on 
graphs. I will come back to this point in the next section. 

The form factor (\ref{K1}) can now be written as double sum over the edge rate 
vectors $\vec{q}$
\be{K2}
K(n,N)  
= \frac{1}{N} \langle \sum_{{\vec{q}, \vec{q}'\in \K_n(G)}} 
A_{\vec{q}} A_{\vec{q}'} P_n(\vec{q}) P_n(\vec{q}')
e^{i ( L_{\vec{q}} - L_{\vec{q}'})}\rangle_{\Delta \tau}\; .
\ee 
A new type of diagonal contribution emerges when considering periodic orbit 
pairs sharing a common $\vec{q}$ -- vector. The total contribution of 
topologically degenerate periodic orbit pairs, which obviously includes the 
original diagonal contributions in the HOdA -- sum rule, is
\be{k-geo}
K_{top}(n,N)= \frac{1}{N}\sum_{\vec{q} \in \K_n(G)}
A_{\vec{q}} ^2 P^2_n(\vec{q}) \sim e^{\alpha_{t} n}
\ee
and $\alpha_t > 0$ in general; (the rate $\alpha_{t}$ can 
be calculated using large deviation techniques (Dembo and Zeitouni
(1993)), strict upper and lower bounds are $0\le \alpha_t \le h_t$, 
and $h_t$ is the topological entropy for  the graph).
All the contributions to $K_{top}$ 
are positive which coincides with a result obtained by Whitney et al (1999)
using diagrammatic techniques for periodic orbit formulae. The diagonal
approximation $K_{top} \sim  \frac{n}{N}$ following from the HOdA -- sum rule 
is valid only for small $\tau = \frac{n}{N}$ when cyclic permutation is
the main source of degeneracies. (This is in general the case for those 
$n$ values for which the majority of orbits visits a given edge at most once).

Unitarity of the underlying $\vec{T}$ matrix implies the asymptotic
result $\lim_{\tau\to\infty} K(\tau,N) = 1$; the exponentially increasing 
topological contributions $K_{top}$ must therefore be 
counterbalanced by additional correlations in the periodic orbit length
spectrum. We will show that these kind of correlations originate from 
the unitarity of the $\vec{T}$ matrix and that the cancelation mechanism is 
extremely sensitive leaving little space for approximate or asymptotic 
treatments.

All what has been said so far is true for arbitrary unitary matrices, 
and thus especially for transfer matrices of quantum graphs and also
for general quantum maps. In order to study the phenomena of periodic orbit 
correlations due to unitarity more closely, I will now focus on a special 
class of chaotic graphs with uniform transition probabilities for which 
all relevant periodic orbit correlations can be given explicitly.

\subsection{Binary graphs and periodic orbit correlations}
\label{sec:sec2.2}
One of the simplest, non-trivial class of graphs are balanced, 
directed binary graphs $B_N$;
these are connected graphs with $N$ vertices ($N$ even)
for which each vertex has exactly two incoming and two outgoing 
edges. The adjacency matrix $\vec{A}_N$ of a binary graph can be written in 
the form
\be{adj_bin} a_{ij} = \left\{ \begin{array}{ll} 
     \delta_{2i,j} + \delta_{2i+1,j} &\mbox{for} \;\; 0 \le i < \frac{N}{2}\\
     \delta_{2i-N,j} + \delta_{2i+1-N,j} &\mbox{for} \; \; \frac{N}{2} \le i < N
                    \end{array} \right . \quad i = 0, \ldots, N-1\; 
\ee
and the number of edges of $B_N$ is $M = 2N$.
Some examples together with their adjacency matrices are 
shown in Figs.\ \ref{Fig:graph2}, \ref{Fig:graph4}, and \ref{Fig:graph6}. 
It will sometimes be useful to switch from a vertex code to an edge code.
A suitable choice is to assign each edge $ij$ corresponding to a non--zero 
matrix element of the adjacency matrix (\ref{adj_bin}) an edge code 
\be{e-code}
i_e = 2 i + j \, \mbox{mod} 2, \quad i_e = 0,1 \ldots, 2 N-1 
\ee
The edge code is given for the examples Figs.\ \ref{Fig:graph2}, 
\ref{Fig:graph4}, and \ref{Fig:graph6}.

Transfer matrices of binary graphs have been studied in connection with 
combinatorial problems for binary sequences (Stanley (1999)), as well as the 
semiclassical quantisation of the Anisotropic Kepler Problem using 
binary symbolic dynamics (Gutzwiller (1988), Tanner and Wintgen (1995)) and 
have been discussed in the context of general quantum maps (Bogomolny (1992)). 
Saraceno (1999)
recently proposed a  quantisation scheme for the baker map which also
leads to quantum maps of the form (\ref{adj_bin}). 

Binary graphs with adjacency matrices (\ref{adj_bin}) are connected, i.e.,
each vertex can be reached from every other vertex, here after at least  
$[\log_2 N] +1$ steps. The topological entropy $h_t = \log2$ independent 
of the order of the graph. The subset of binary graphs of order 
$N = 2^k, k\in \N$, the so--called {\em de Bruijn} -- graphs (Stanley (1999)), 
deserves special attention; the dynamics on these graphs can directly be 
related to the set of all binary sequences and there exists a one-to-one 
relation between finite binary symbol strings ($a_1, a_2, \ldots a_n$), 
$a_i \in \{0,1\}$ of length $n$ and the periodic orbits of the graph i.e.\
\[
(a_1, a_2, \ldots a_n) \leftrightarrow (v_1, v_2, \ldots v_n) 
\quad a_i \in \{0,1\}, \; v_i \in \{0,2^k-1\}
\]
with
\[ 
v_i = \sum_{j=1}^{k} a_{i+j-1} 2^{k-j}  \quad \mbox{and} \; \;
a_{i+n} = a_i \; 
\]
for graphs of order $N = 2^k$. The number of orbits of period $n$ on 
these graphs is exactly $2^n$. \\

\begin{figure}
\centering
\centerline{
         \epsfxsize=4.5cm
         \epsfbox{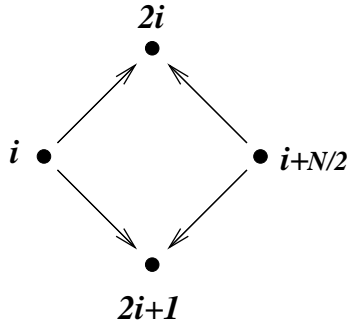}
         }
\caption[]{\small
Local network of correlated edge lengths, see Eq.\ (\ref{uni_cond_II}); 
opposite edges form a pair, the two pairs have a combined length 
difference of $\pi$.
}
\label{Fig:local}
\end{figure}

I will consider unitary transfer matrices of binary graphs next.
The unitarity conditions for a transfer matrix $\vec{T}_N$ of a binary 
graph with adjacency matrix (\ref{adj_bin}) can be stated simply in terms
of the unitarity conditions for the set of $N/2$ different $2 \times 2$ 
matrices $\vec{u}_i$ with
\be{uni_cond_I} \vec{u}_i = \left( \begin{array}{ll} 
                            t_{i,2i} &t_{i,2i+1} \\
                            t_{i+\frac{N}{2},2i} & t_{i+\frac{N}{2},2i+1}
                            \end{array} \right) \qquad i = 1,2,\ldots, N/2 \; .
\ee
Focusing on unitary binary transfer matrices with uniform local spreading, 
i.e., setting $|t_{ij}| = 1/\sqrt{2}$ for all non-zero matrix elements of 
$\vec{T}_N$, the unitarity condition can be written 
as a pure phase correlation. One obtains the following 
relation between the lengths of edges
\be{uni_cond_II} 
\left(L_{i,2i} + L_{i+\frac{N}{2},2i+1}\right) - 
\left(L_{i,2i+1} + L_{i+\frac{N}{2},2i}\right) \bmod 2 \pi = \pi  
\qquad i = 0,2,\ldots, N/2-1 \; ,
\ee
the corresponding local network is shown in Fig.\ \ref{Fig:local}. 
The unitary condition (\ref{uni_cond_II}) will be shown to be responsible 
for the periodic orbit correlations relevant to balance out the exponentially 
increasing topological contributions to the form factor $K(\tau,N)$.
Its simplicity makes it possible to turn the problem of finding periodic 
orbit length correlations into a combinatorial problem of finding all 
exact periodic orbit degeneracies (up to phase differences being a multiple
of $\pi$), which can be done in principle. 

The dynamics described by the classical transfer matrix with
transition probabilities $t^{cl}_{ij} = 1/2$ for all possible transitions 
$ij$ in the binary graph is maximally mixing for the geometry 
(\ref{adj_bin}), i.e.\ $h_t = K  = \log2$ and $K$ is the Kolmogorov 
entropy for graphs (Schuster (1989)); the 
conditions (a) -- (c) in section \ref{sec:sec1} are thus satisfied 
as long as there are no systematic edge length correlations present except  
from those introduced through Eq.\ (\ref{uni_cond_II}). 
One can furthermore show that the generalised diagonal contribution 
(\ref{k-geo}) increases exponentially with a rate $\alpha_t = h_t = \log 2$  
independent of the order of the binary graph.\\

Periodic orbit correlations beyond topological degeneracies can
be expressed in terms of edge and vertex staying rates. The vertex
staying rates $\tilde{q}_{i}(\v)$ of an orbit $\v$ are defined 
analogue to (\ref{edge_stay}) as the number of times a periodic orbit
visits a vertex $i$, i.e.
\be{vertex_stay} 
\tilde{q}_{i}(\v) = \sum_{l=1,n} \delta_{i,v_l} \qquad i\in V(G)\;  .
\ee
Vertex and edge staying rates are connected by conservation laws 
(or shift invariance properties (Dembo and Zeitouni (1993)))
of the form
\begin{eqnarray} \label{cons-q}
q_{i,2i} + q_{i,2i+1} &=& q_{[\frac{i}{2}],i} +
q_{[\frac{i}{2}]+\frac{N}{2},i} = \tilde{q}_i \qquad
\forall i=0,\ldots,\frac{N}{2} -1\\
\underbrace{q_{i,2i-\frac{N}{2}} + q_{i,2i-\frac{N}{2}+1}}_{\mbox{incoming 
edges}}
&=& 
\underbrace{q_{[\frac{i}{2}],i} + q_{[\frac{i}{2}]+\frac{N}{2},i}}_{
\mbox{outgoing edges}}
= \tilde{q}_i \qquad \forall i=\frac{N}{2},\ldots,N-1 \nonumber
\end{eqnarray}
and $[.]$ denotes the integer part. 
A direct consequence of (\ref{uni_cond_II}) and (\ref{cons-q}) 
is the following condition for periodic orbit correlations:\\

{\em All periodic orbits having the same vertex staying rates $\tilde{\vec{q}} =
(\tilde{q}_0, \ldots, \tilde{q}_{N-1})$ differ in length exactly by a 
multiple of $\pi$.}\\

This can be shown by noting that for two orbits $\v, \v' \in 
{\cal PO}_n(B_N)$
with 
$\Delta \tilde{\vec{q}} = \tilde{\vec{q}}(\v) - \tilde{\vec{q}}(\v') =\vec{0}$, 
one obtains 
\[ \Delta q_{i,2i} + \Delta q_{i,2i+1} = 0, \quad 
\Delta q_{i,2i} + \Delta q_{i+\frac{N}{2},2i} = 0, \quad 
\Delta q_{i+\frac{N}{2},2i+1} + \Delta q_{i+\frac{N}{2},2i} = 0 \; ,
\]
see also Fig.\ \ref{Fig:local}. One therefore has
\[\Delta q_{i,2i} = \Delta q_{i+\frac{N}{2},2i+1} = 
-\Delta q_{i,2i+1} = - \Delta q_{i+\frac{N}{2},2i} \]
which together with (\ref{uni_cond_II}) yields 
\be{length-d}
\Delta L = L_{\v}- L_{\v'}  = \pi \sum_{i=0}^{N/2-1} \Delta q_{i,2i}
= \pi d_{\v,\v'} \; .
\ee
The corresponding contribution of the periodic orbit pair to the form factor 
(\ref{K1}) is then $(-1)^{d_{\v,\v'}} 2^{-n}$. Note that 
the amplitudes $A_{\v}$ equal $2^{-n/2}$ for all orbits of period $n$. 

The from factor can thus be written as a sum over weighted correlations of
the degeneracy function (\ref{deg-func}), i.e.\ 
\begin{eqnarray}\label{K-N}
K(n,N) &=& 
\frac{1}{N}\frac{1}{2^{n}}
\sum_{\tilde{\vec{q}}\in \tK_n(B_N)} 
\left(
\sum_{\vec{q}}\sum_{\vec{q}'} (-1)^{\sum_{i} \Delta q_{i,2i}} 
P_n(\vec{q}) P_n(\vec{q'}) \right)  \nonumber \\
&=&
\frac{1}{N}\frac{1}{2^{n}}
\sum_{\tilde{\vec{q}}\in \tK_n(B_N)} 
\left(
\sum_{\vec{q}} (-1)^{[\sum_{i}q_{i,2i}]} 
P_n(\vec{q})\right)^2 \, 
\end{eqnarray}
and $[.]$ denotes the integer part. The sum is taken over the 
$N$ dimensional integer lattice $\tK_n(B_N)$
of possible vertex staying rate vectors $\tilde{\vec{q}}$ corresponding
to periodic orbits of period $n$ of a binary graph $B_N$; the vectors
$\vec{q}$, $\vec{q'}$  correspond here to the $N/2$ components 
$(q_{i,2i},\; i=0, N/2-1)$ of the total edge staying rate vector only. 
The contributions of periodic orbit pairs which are not correlated
by having length differences being a multiple of $\pi$ are assumed
to be Gaussian distributed with zero mean. We will neglect these 
random background contributions from now on and concentrate on the 
contributions from correlated periodic orbit pairs only.\\

Before turning to the problem of calculating degeneracy functions, 
a few remarks on the edge staying rates.
The components of the edge staying rate vector $\vec{q}$ are related to each
other by the shift invariance properties (\ref{cons-q}).
These are $N$ conditions which can be shown to lead to $N-1$ independent 
equations for the $2 N$ rates $q_{ij}$; together with the restriction 
\be{part-cons}
\sum_{i=0}^{N-1} \tilde{q}_i = n
\ee
for orbits of period $n$, one can write the edge staying rates in terms of
$N$ independent quantities, which effectively allows to half the dimension of 
$\K_n(B_N)$. The length degeneracy functions $P_n$ depends thus on $N$ 
independent variables only.

There are further restrictions on the independent components of $\vec{q}$.
Apart from the obvious condition $q_{ij} \ge 0$ $\forall ij \in E(B_N)$, one
must also ensure that
the sum over the $N$ independent components of $\vec{q}$ does not exceed 
$n$ and that the staying rates do correspond to a connected, closed path on 
the graph. An example for an edge staying rate vector $\vec{q}$ violating the 
last restriction is $\vec{q} = (q_{00}, 0, \ldots, 0, q_{N-1,N-1})$ with 
$q_{00} \ne 0$ and $q_{N-1,N-1} \ne 0$ which corresponds to two
disconnected periodic orbits. I will come back to the problem of determining 
the lattice $\K_n$ in more detail in the next section.

\section{Periodic orbit length degeneracy functions -- analytic results}
\label{sec:sec3} 
The periodic orbit length correlations in binary graphs with constant 
transition amplitudes can be completely described in terms of the 
degeneracy function (\ref{deg-func}). The problem of calculating 
the form factor is thus converted to a combinatorial problem of finding the
number of closed (connected) paths on a graph which visit each edge
the same number of times. This problem can be treated explicitly for 
low-dimensional graphs; results for binary graphs up to order 6 will be
presented here. 
\subsection{Binary graphs of order $N = 2$}
\label{sec:3.1}
The case $N = 2$ has already been treated by Schanz and Smilansky (1999) 
in somewhat different circumstances. 
\footnote{Schanz and Smilansky (1999) analysed unitary $2\times 2$ matrices in 
connection with simple quantum (star--) graphs. The unitary transfer matrices 
considered have the extra constraint $L_{01} = L_{10}$. It can, however, 
be shown that this conditions does not lead to additional periodic orbit 
length correlations, see also Sec.~\ref{sec:sec4}.}
We will re-derive some of the results in order to introduce the basic 
notations and concepts which will be useful when considering binary graphs
for $N > 2$. Some new asymptotic results for the two-dimensional case will 
also be presented here.

\begin{figure}
\centering
\centerline{
         \epsfxsize=9.0cm
         \epsfbox{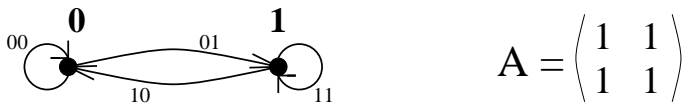}
         }
\caption[]{\small
Binary graph of order 2 together with its adjacency matrix $\vec{A}$.
}
\label{Fig:graph2}
\end{figure}
A binary graph of order 2 is shown in Fig.~\ref{Fig:graph2}. The shift 
invariance property, Eq.~(\ref{cons-q}), implies the following conditions for
the edge staying rate vector $ \vec{q} = (q_{00},q_{01},q_{10},q_{11})$, i.e.
\begin{eqnarray} \label{cons-q-2}
\tilde{q}_0 &=& q_{00} + q_{01} = q_{10}+q_{00}, \\\nonumber
\tilde{q_1} &=& q_{11} + q_{10} = q_{01}+q_{11}
\end{eqnarray}
and $\tilde{q}_{0}$, $\tilde{q}_{1}$ represent the vertex staying rates.
After choosing $q_{00}$ and $q_{11}$ as independent variables and together 
with the condition (\ref{part-cons}), one obtains
\be{q-cond-2} 
q_{01} = q_{10} = \frac{1}{2}(n-q_{00}-q_{11}), \quad 
\tilde{q}_{0} = \frac{1}{2}(n+q_{00}-q_{11}), \quad 
\tilde{q}_{1} = \frac{1}{2}(n-q_{00}+q_{11})\; ,
\ee
for orbits of period $n$.

\begin{figure}
\centering
\centerline{
         \epsfxsize=10.0cm
         \epsfbox{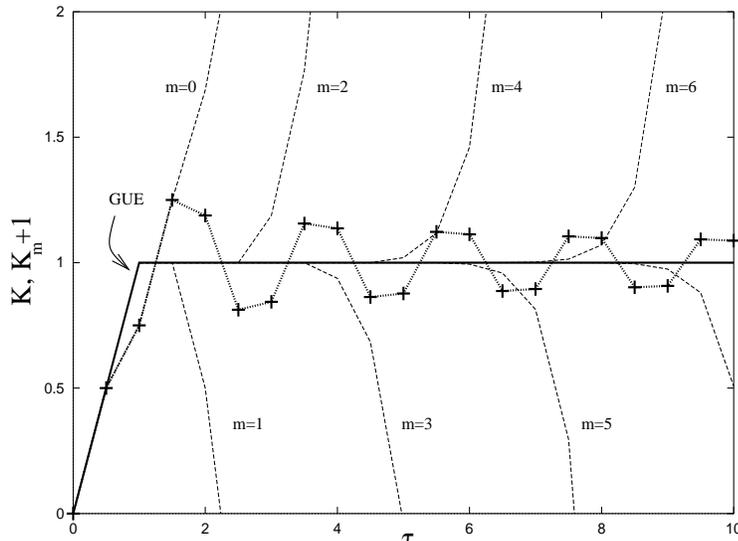}
         }
\caption[]{\small The form factor for binary graphs of order 2 (dotted line
with $+$) is shown as function of $\tau = \frac{n}{2}$; the partial sums 
$K_m$ contribute with alternating signs starting at $\tau = m+1$; 
(the dashed lines correspond to $K_m +1$ for $m > 0$). The GUE - form factor 
is also displayed for comparison.

}
\label{Fig:k-2}
\end{figure}

The periodic orbit length degeneracy function $P_n(q_{00},q_{11})$ can 
be derived by starting with the special case $q_{00}=q_{11}=0$. 
One immediately obtains $P_n(0,0) = 2$ for $n$ even; the two
periodic orbits correspond to the $\frac{n}{2}$--th repetition of the 
primitive periodic orbits $01$ and $10$ of period $2$. 
It is advantageous to switch to an edge symbol code, i.e., to identify
\[ 00 \to 0_e; \quad 01 \to 1_e \quad 10 \to 2_e; \quad 11 \to 3_e\; ,\]
see also Eq.\ (\ref{e-code} and Fig.~\ref{Fig:graph2}. The two orbits 
$01$ and $10$ can then be written as
\be{po-2} 
\underbrace{1_e2_e1_e2_e\ldots1_e2_e}_{n}, 
\quad\mbox{and} \quad  
\underbrace{2_e1_e2_e1_e\ldots2_e1_e}_{n}.
\ee
The symbol $0_e$ can only occur after the symbol $2_e$ and it can be repeated.
A periodic orbit of period $n+q_{00}$ can thus be obtained by inserting 
$q_{00}$ symbols $0_e$ in between the $2_e1_2$ blocks in the periodic orbit 
sequences (\ref{po-2}). Symbols $0_e$ can be placed at $\frac{n}{2}+1$ 
positions for the first orbit in (\ref{po-2}) and $\frac{n}{2}$ positions 
for the second orbit. Similar arguments apply for inserting $q_{11}$ 
symbols $3_e$ into the sequences (\ref{po-2}). Using standard combinatorial 
formulae to find the number of combinations to distribute $q_{00}$ 
items among $\frac{n}{2}+1$ or $\frac{n}{2}$ boxes with repetitions,
one obtains
\[ 
P_{n+q_{00}+q_{11}} (q_{00},q_{11}) = 
\bino{\frac{n}{2}+q_{00}}{q_{00}} \bino{\frac{n}{2}+q_{11}-1}{q_{11}} 
+ \bino{\frac{n}{2}+q_{00}-1}{q_{00}} \bino{\frac{n}{2}+q_{11}}{q_{11}} \; .
\]
After rescaling ($n+q_{00}+q_{11}$) to $n$ and using the relations 
(\ref{q-cond-2}) one may write the degeneracy function as
\be{def-fun-2}
P_{n} (q_{00},q_{11}) = 
\frac{n}{q_{01}}\bino{\tilde{q}_{0}-1}{q_{00}} 
\bino{\tilde{q}_{1}-1}{q_{11}} \; .
\ee
The possible integer values for $q_{00}$ and $q_{11}$ have to obey certain 
restrictions which follow directly from (\ref{q-cond-2}), i.e.\
\be{e-2-cond} 
q_{00} + q_{11} < n \quad 
\mbox{and} \quad (n - q_{00} - q_{11}) \bmod 2 = 0\, . 
\ee
The degeneracy function (\ref{def-fun-2}) approaches a Gaussian distribution
in the limit $n\to\infty$; its form can be derived with the help of
large deviation techniques (Dembo and Zeitouni (1993)), i.e., one obtains
\begin{eqnarray} \label{def-fun-2-app}
&&P_{n} (q_{00},q_{11}) \sim 
\frac{4}{\pi n } 2^{n} e^{-n ( 4x^2 + y^2)} \\ \nonumber
&&\mbox{with}  \qquad 
x = \frac{1}{n \sqrt{2}} (q_{00} + q_{11} - \frac{n}{2}), \qquad
y = \frac{1}{n \sqrt{2}} (q_{00} - q_{11}) \, .
\end{eqnarray}
The asymptotic result (\ref{def-fun-2-app}) is too crude to be useful in a
calculation of the form factor directly; it does provide, however, some insight
into the asymptotic behaviour of the various contributions entering the form 
factor.  Especially the contributions of topologically degenerate periodic 
orbit pairs, see Eq.~(\ref{k-geo}), can be estimated to be 
\[ 
K_{top}(n) \sim \frac{1}{2^{n+1}}
\int \int dq_{00} dq_{11} \, P^2_{n}(q_{00},q_{11}) = \frac{2^n}{\pi n}\, ,
\]
and one obtains $\alpha_{t} = \log 2$ for the growth rate of the diagonal
contributions (\ref{k-geo}). Periodic orbit pairs being
degenerate up to a phase difference $m \pi$ enter the form factor 
asymptotically as 
\[ 
K_{m}(n) \sim \frac{(-1)^m}{2^{n+1}}
\int \int dq_{00} dq_{11} \, P_{n}(q_{00},q_{11}) P_{n}(q_{00}+m,q_{11}+m) 
= (-1)^m \frac{2^n}{\pi n} e^{-4 \frac{m^2}{n}}\, .
\]
The form factor thus consists of an increasing number of exponentially 
growing terms which differ in sign (see also Fig.~\ref{Fig:k-2}). Only a very 
delicate balance between these terms ensures the cancelations necessary to
lead to the asymptotic behaviour  $\lim_{n\to\infty}K(n) = 1$. The 
approximations 
above are indeed not sufficient to preserve the asymptotic limit and
give exponentially growing terms for large $n$; similar
arguments might hold for the breakdown of semiclassical approximations to 
quantum form factors, see e.g.\ Tanner (1999). Note also, that the 
diagonal terms relevant for the HOdA -- sum rule do not play a prominent role 
in the discussion above; they give a linear contribution to $K_{top}$ which 
is already sub-dominant for moderate $n$ values.

The periodic orbit pair contributions to the form factor can be computed
explicitly by summing the exact length degeneracy function (\ref{def-fun-2})
over the possible edge staying rates obtained from the conditions 
(\ref{e-2-cond}). It may be written in compact form in the following way
(Schanz and Smilansky (1999))
\be{K-2}
K(n) = 
\frac{1}{2^{n+1}}\left[2 + 
\sum_{\tilde{q}_0= 1}^{n-1} 
\left(\sum_{q_{01}=1}^{\frac{n}{2} - |\frac{n}{2}-\tilde{q}_0|)} 
(-1)^{q_{01}} P_n(q_{00},q_{11})\right)^2\right] = 
1 + \frac{(-1)^{n+l}}{2^{2l+1}} \bino{2l}{l}  ,
\ee
with $l = [n/2]$ and $\tilde{q}_0$, $q_{01}$ can be expressed in terms of 
$q_{00}, q_{11}$ using (\ref{q-cond-2}).
It is a remarkable fact that the sum can be determined explicitly, a result 
derived by Schanz and Smilansky (1999) using quantum graph techniques. 
The form factor, Eq.~(\ref{K-2}), is displayed in Fig.~\ref{Fig:k-2}
together with the asymptotic results. $K(n)$ approaches 1 in the large $n$ 
limit, but is different from the RMT-result for $2 \times 2$ matrices. The 
periodic structure can be seen to coincide with the start of a new family of 
degenerate orbits and is thus a remnant of non--perfect cancelation of the
various $K_m(n)$ contributions. Convergence of the correlated periodic orbit 
pair contributions to the  RMT -- result is observed when 
increasing the order $N$ of the binary graph as will be shown in the next 
sections.

\subsection{Binary graphs of order $N = 4$ and $N=6$}
\begin{figure}
\centering
\centerline{
         \epsfxsize=10.0cm
         \epsfbox{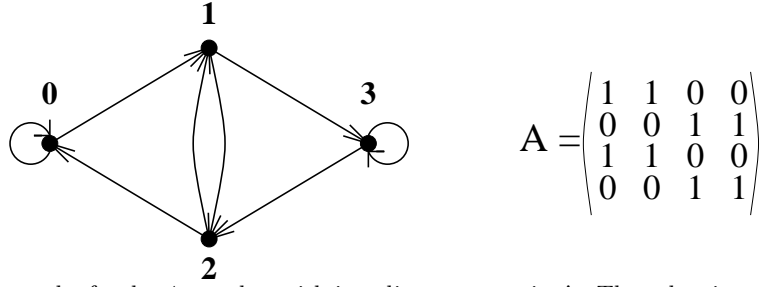}
         }
\caption[]{\small
Binary graph of order 4 together with its adjacency matrix $\vec{A}$. The
subscript $e$ denotes the edge symbol code.
}
\label{Fig:graph4}
\end{figure}

The edge and vertex staying rates of periodic orbits of a binary graph of 
order $N = 4$, see Fig.\ \ref{Fig:graph4}, can be written in terms of 4 
independent variables. A possible choice for the edge staying rates is
$q_{00}$, $q_{12}$, $q_{21}$ and $q_{33}$. The other edge and 
vertex rates can be computed by using Eqs.~(\ref{cons-q}), explicit formulae 
are given in appendix \ref{app:A}. 

The periodic orbit length degeneracy function can be obtained by 
arguments similar to the one described in the last section. The discussion is 
somewhat technical and is referred to appendix \ref{app:A}.  
The final result is
\be{def-fun-4}
P_{n} (q_{00},q_{12},q_{21},q_{33}) = 
\frac{n}{\tilde{q}_{1}}
\bino{\tilde{q}_{1}}{q_{01}} \bino{\tilde{q}_{2}}{q_{21}} 
\bino{\tilde{q}_{0}-1}{q_{00}} \bino{\tilde{q}_{3}-1}{q_{33}} \; ,
\ee
and $\tilde{q_i}$ denotes again the vertex staying rates. The possible
entries on the 4--dimensional integer $\vec{q}$ lattice can be stated by
conditions similar to those in Eq.~(\ref{e-2-cond}). Periodic orbits which 
differ in length by a multiple of $\pi$ have the same vertex staying rates 
but may differ in the variables
\be{tu} s_0 = q_{00} + q_{21}, \quad s_1 =  q_{12} + q_{33}.\ee
The length difference for orbits with identical vertex rates
is given by $\Delta L = \frac{1}{2}(\Delta s_0 + \Delta s_1) \pi$,
see Eq.\ (\ref{length-d}). The form factor can be written in 
terms of degenerate periodic orbit pairs only and one obtains
\be{K-4}
K(n) = 
\frac{1}{4}\frac{1}{2^{n}}\left(2 + 
\sum_{\tilde{q}_0 + \tilde{q}_1= 1}^{n-1} 
\sum_{\tilde{q}_{1}=1}^{\frac{n}{2}- |\frac{n}{2}-\tilde{q}_0-\tilde{q}_1|)} 
\left(\sum_{s_0=|q_{00} - q_{21}|, 
s_1=|q_{12} - q_{33}|}^{s_0+s_1 < n}
(-1)^{[\frac{s_0+s_1}{2}]} P_n(\vec{q})\right)^2 \right)\, .
\ee

\begin{figure}
\centering
\centerline{
         \epsfxsize=13.0cm
         \epsfbox{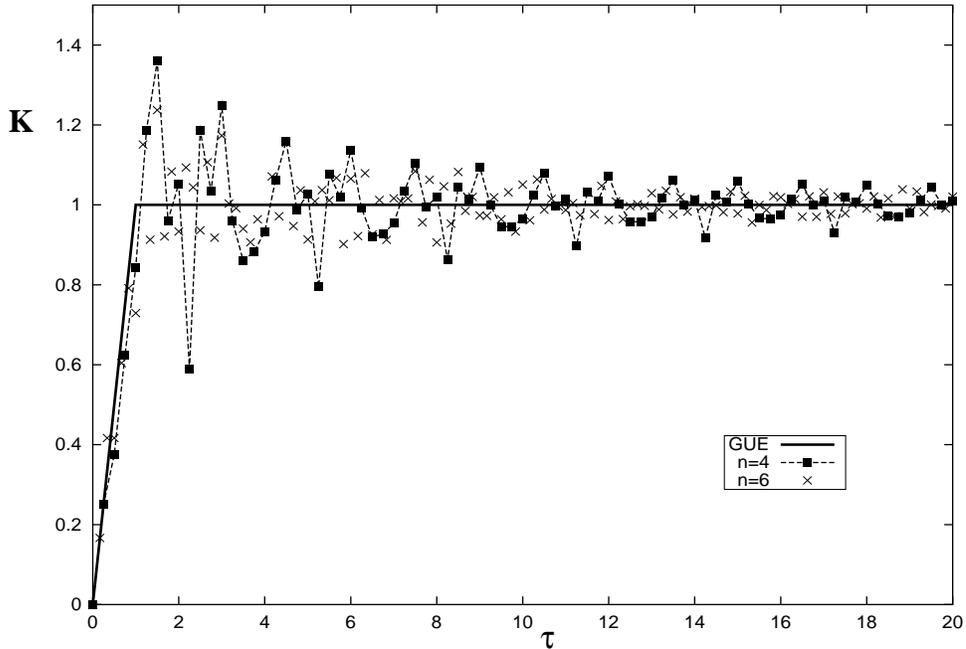}
         }
\caption[]{\small
Periodic orbit pair contributions to the form factor for binary graphs of 
order $N = 4$ and $N=6$.
}
\label{Fig:k-ex}
\end{figure}

The form factor $K(\tau)$ with $\tau = n/4$ obtained from Eq.\ (\ref{K-4}) 
is shown in Fig. \ref{Fig:k-ex}. It oscillates periodically with decreasing 
amplitude about the RMT - result similar to the behaviour observed in the 
case $N=2$, see Fig.\ \ref{Fig:k-2}. A closed expression for the sum similar
to Eq.\ (\ref{K-4}) could not be found.

The sums in (\ref{K-4}) are already quite cumbersome and the number of 
summation 
variables increases with the order $N$. The number and complexity of 
the restrictions for the $\vec{q}$ -- lattice $\K_n(B_N)$ increases
accordingly. The case $N=6$ can, however, still be treated along 
the ideas developed above; it will be presented here as a last example 
for obtaining the form factor by summing  over the periodic orbit length 
degeneracy function. \\

\begin{figure}
\centering
\centerline{
         \epsfxsize=12.0cm
         \epsfbox{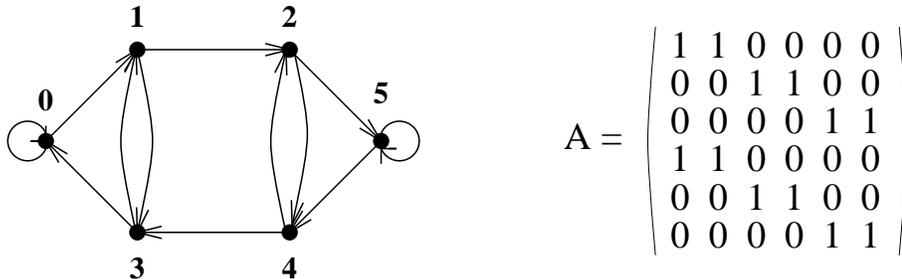}
         }
\caption[]{\small
Binary graph of order 6 together with its adjacency matrix $\vec{A}$.
The subscript $e$ denotes the edge symbol code.
}
\label{Fig:graph6}
\end{figure}
The binary graph of order $N=6$ is shown in Fig.\ \ref{Fig:graph6}. 
A possible choice for the independent edge staying rates is 
$q_{00}$, $q_{13}$, $q_{24}$, $q_{31}$, $q_{42}$, and $q_{55} $.
The derivation of the degeneracy  function can again be found in 
appendix \ref{app:A}, the final result is
\be{def-fun-6}
P_{n} (\vec{q}) = 
\frac{n q_{12}}{\tilde{q}_{2} \tilde{q}_{3}}
\bino{\tilde{q}_{1}}{q_{31}} \bino{\tilde{q}_{2}}{q_{42}} 
\bino{\tilde{q}_{3}}{q_{13}} \bino{\tilde{q}_{4}}{q_{24}} 
\bino{\tilde{q}_{0}-1}{q_{00}} \bino{\tilde{q}_{5}-1}{q_{55}} \; .
\ee
The vertex rates $\tilde{q_i}$ and the edge rate $q_{12}$
entering (\ref{def-fun-6}) can be expressed in terms of the independent
variables $\vec{q} = (q_{00}, q_{13}, q_{24}, q_{31}, q_{42}, q_{55})$, 
see appendix \ref{app:A}. The summation over the six dimensional 
lattice $\K_n(B_6)$ of possible $\vec{q}$ vectors can be stated in 
terms of the vertex staying rates and the variables
\[ s_0 = q_{00} + q_{31}, \quad s_1 = q_{12} + q_{43}, \quad 
s_2 = q_{24} + q_{55} \; .\]
The expression for the form factor as sum over degenerate periodic
orbit pairs is thus
\be{K-6}
K(n) =
\frac{1}{6}\frac{1}{2^{n}}
\sum_{\tilde{q}_0= 1}^{n}
\sum_{\tilde{q}_{1}=0}^{[(n-\tilde{q}_0)/2]}
\sum_{\tilde{q}_{2}=0}^{[(n-\tilde{q}_0-2\tilde{q}_1)/2]}
\left(\sum_{s_0,s_1,s_2}^{(s_0+s_1+s_2)<n}
(-1)^{[\frac{s_0+s_1+s_2}{2}]} P_n(\vec{q})\right)^2 \,
\ee
and the inner sum runs over all possible $s_i$, $i=0,1,2$ values.
The form factor $K(\tau)$ after summing Eq.\ (\ref{K-6}) is
displayed in Fig. \ref{Fig:k-ex} with $ \tau= n/6$.
The sums, Eqs.~(\ref{K-4}) and (\ref{K-6}), follow the RMT -- result
more closely than in the $N = 2$ case, see Fig.\ \ref{Fig:k-2}. The linear 
behaviour for $\tau < 1$ starts to emerge and convergence to the asymptotic 
result $K\to1$ is observed in the large $\tau = n/N$ limit. 

Larger matrices have to be considered in 
order to test convergence of degenerate periodic orbit pair contributions 
towards the RMT form factor for all $\tau$. Determining the degeneracy function
and the lattice conditions $\K_n(B_N)$ becomes increasingly difficult for 
graphs of order $N > 6$. In the next section, I will therefore present 
results obtained from counting all correlated periodic orbit pair 
contributions directly.

\section{Periodic orbit pair contributions to the from factor for 
de Bruijn Graphs of order $N \ge 8$.}
\label{sec:sec4} 
The periodic orbit pair contributions to the form factor can be 
calculated directly by determining the set of periodic orbits of given
period $n$ and calculating periodic orbit degeneracies with the help 
of edge and vertex staying rates and the condition 
(\ref{length-d}). The task of finding the set of periodic orbits is 
especially simple for de Bruijn graphs, i.e.\ for binary graphs of order 
$N = 2^r$, due to the one-to-one relation between periodic orbits and 
finite binary symbol strings, see section \ref{sec:sec2.2}.

Counting the periodic orbit pair degeneracies explicitly does, however,
seriously limit the range of periods over which periodic orbit correlations
can be considered. Due to the exponential increase in the number of orbits
only values up to $n \approx 26$ could be reached. This in turn sets 
an upper bound on the $\tau = n/N$ values for which the form factor can be 
studied. \\

\begin{figure}
\centering
\centerline{
         \epsfxsize=12.0cm
         \epsfbox{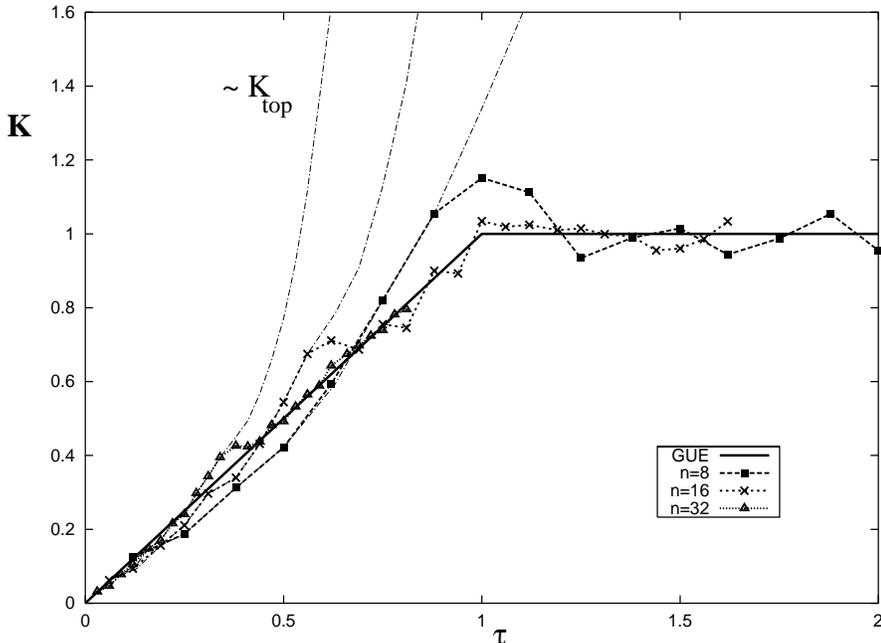}
         }
\caption[]{\small
Periodic orbit pair contributions to the form factor for binary graphs of 
order $N = 8, 16$ and $32$. The small $\tau$ behaviour is dominated
by exponentially increasing topological contributions (dashed lines).
}
\label{Fig:k-c}
\end{figure}

\noindent
\underline{GUE -- results}:\\
Results for $N = 8$, 16 and 32 and no further symmetry present are shown in 
Fig.\ \ref{Fig:k-c}. One observes a convergence of the periodic orbit 
pair contributions to the GUE-result; the kink at $\tau = 1$ is resolved for 
binary graphs of order $N = 16$; the periodic orbit results follows the linear
behaviour for $\tau < 1$ even closer for $N = 32$. It was 
not possible to extend the results for $n = 32$ to the critical time 
$\tau = 1$ due to the restrictions on the available $n$ values.
The small $\tau$ behaviour is dominated by the exponentially increasing
topological contributions, see Fig.\ \ref{Fig:k-c}. The so--called diagonal 
contributions due to cyclic permutations of periodic orbit codes are 
important only in the small $\tau$ regime, i.e.\ $\tau \sim \log_2(N)/N$, 
before vertex exchange degeneracies set in. \\

\noindent
\underline{GOE -- results}:\\
So far only unitary transfer matrices without symmetries have
been considered. Symmetries in the dynamics impose additional correlations 
on periodic orbit length spectra and do have an effect on the spectral 
statistics. Time reversal symmetry is of special importance as it occurs 
frequently in physical systems; correlations due to time reversal symmetry
are in addition non-trivial leading to a form factor which is not piecewise
linear as in the GUE case; only the 
linear behaviour for $K(\tau)$ in the limits $\tau \to 0$ and 
$\tau \to \infty$ is understood in terms of semiclassical 
arguments (Berry (1985)).

It is a priori not clear how to establish time reversal symmetry 
for the dynamics on an arbitrary directed graph. Time reversal symmetry
can, however, be constructed for de Bruijn graphs of order $N=2^k$ using 
the underlying binary symbolic dynamics and the edge code, 
Eq.\ (\ref{e-code}). The edge code can be written in terms of a 
binary string of length $k+1$ such that
$i_e = \sum_{l=1}^{k+1} a_l(i_e) 2^{k+1-l}$, and 
$\vec{a}(i_e) = (a_1,\ldots a_{k+1})$ is a binary string of length $k+1$ 
with $a_l \in {0,1}$. Time reversal symmetry can be established by imposing
\be{uni_cond_III}  
L_{i_e} = L_{i_e'},\;\; r_{i_e} = r_{i_e'} \qquad \mbox{if} \qquad 
\vec{a}(i_e) = 
\overline{\vec{a}}(i_e')
\ee
for the edge lengths and transition rates  
and $\overline{\vec{a}}$ denotes the code $\vec{a}$ written backwards,
i.e.\ $i_e' = \sum_{l=1}^{k+1} a_l(i_e) 2^{l-1}$. The condition, 
Eq.\ (\ref{uni_cond_III}), and the unitarity condition, Eq.\ 
(\ref{uni_cond_II}), are assumed to be the only sources of correlations in 
the periodic length spectrum.

Time reversal symmetry does not effect graphs of the order $N \le 8$, 
$ N = 2^k$. 
This is due to the fact that the edge staying rates for a given edge and 
its time reversed partner are related by conservation laws (\ref{cons-q}) 
such that there are no further degeneracies for these low dimensional cases.
One finds for $N=2$, for example, that $q_{01} = q_{10}$ and a periodic
orbit and its time reversed partner are always in the same degenaracy 
class even if the condition (\ref{uni_cond_III}) is not imposed. 
\footnote {The unitary $2\times 2$ matrices studied by 
Schanz and Smilansky (1999) do thus correspond to time reversal symmetric 
(binary) graphs.} 

\begin{figure}
\centering
\centerline{
         \epsfxsize=12.0cm
         \epsfbox{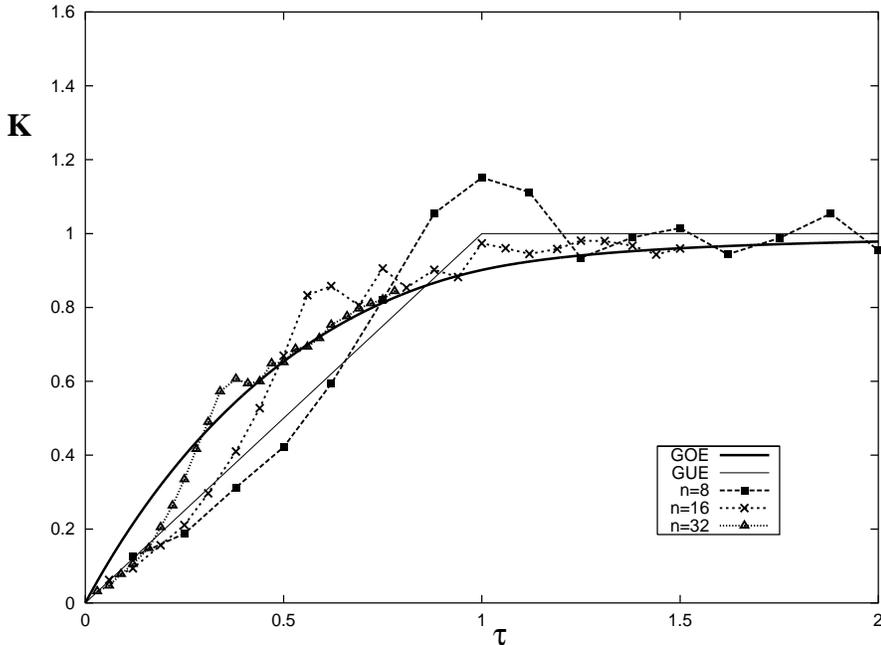}
         }
\caption[]{\small
Periodic orbit pair contributions to the form factor for binary graphs of 
order $N = 8$, $N=16$ and $N = 32$ and time reversal symmetry.
}
\label{Fig:k-co}
\end{figure}

Results for graphs with time reversal symmetry are shown in 
Fig.~\ref{Fig:k-co}; the case $N = 8$ is indeed identical to the non-time 
reversal symmetric
result in Fig.~\ref{Fig:k-c}. The results for $N = 16$ and 32 are, however,
different from those in Fig~\ref{Fig:k-c}; the periodic orbit 
pair contributions approach the GOE - result and not the GUE form factor
with increasing $N$. The condition (\ref{uni_cond_III}) does therefore 
introduces new correlations among periodic orbits for $N > 8$ which are 
beyond the additional topological degeneracy between an orbit and its time 
reversed partner giving rise to a factor 2 in the HOdA -- diagonal 
approximation. Note also the exponentially increasing components for small 
$\tau$ due to topological degeneracies similar to those in Fig.\ 
\ref{Fig:k-co}.

\section{Conclusions}
Degeneracies in the length spectrum of periodic orbits of generic directed 
graphs have been studied. Transition rates and edge lengths in the graph 
are identified with the amplitudes and phases of matrix elements of the 
complex 
transition matrices. General concepts like edge and vertex staying rates
as well as the periodic orbit length degeneracy function have been 
introduced.  The form factor can be written in terms of the degeneracy 
function revealing an exponentially increasing `diagonal contribution'
due to topologically degenerated orbits. Topological degeneracies
exist independently of the actual choice of length segments on the graph 
(defined through the transition matrix) and are a purely 'classical' effect 
depending only on the topology of the graph. Further 
correlations amongst orbits are introduced when considering unitary 
transfer matrices. 

These correlations have been studied for a particular simple class of 
graphs, so--called binary graphs with constant transition amplitudes. 
The correlations can be given explicitly in terms of edge and vertex 
staying rates. One finds in particular that periodic orbits which have the 
same vertex staying rates differ in length by exactly a multiple of $\pi$. 
Finding the periodic orbit degeneracy function turns into a combinatorial 
problem which has been solved for binary graphs with up to 6 vertices. 

The form factor can be shown to consist of exponentially increasing 
contributions which balance in a very delicate way to give 
$\lim_{\tau\to\infty} K(\tau) = 1$. The periodic orbit sums also reveal
convergence towards the RMT - result for intermediate $\tau$ -- values when 
increasing the order of the graph, both for time reversal and non-time
reversal symmetric binary graphs. Binary graphs may thus turn out to
be an ideal model systems to study the connection between periodic orbit 
formulae and random matrix theory. Eigen spectra of
binary graphs follow generic random matrix statistics in the large $N$ limit 
and periodic orbit correlations are known explicitly. \\[.5cm]

\noindent
{\large \bf Acknowledgments}\\

\noindent
Parts of the work has been carried out at BRIMS, Hewlett--Packard
Laboratories, Bristol; I would like to thank Jeremy Gunawardena for
the hospitality experienced throughout my stays, Neil O'Connell for 
stimulating discussions and Gregory Berkolaiko for comments on the 
manuscript.


References


\begin{appendix}
\section{Periodic orbit length degeneracy function for binary graphs: 
Exact results}
\label{app:A}
The expressions for periodic orbit length degeneracy functions for 
binary graphs of order $N = 4$ and $N =6$, Eqs.\ (\ref{def-fun-4}) and 
(\ref{def-fun-6}), will be derived here.\\

\noindent
\underline{The case $N=4$}:\\[.3cm]
As for binary graphs of order $N = 2$ discussed in section \ref{sec:3.1},  
it is useful to switch to an edge symbol code, see Eq.\ (\ref{e-code}); 
adopting the vertex
symbol code of Fig.\ \ref{Fig:graph4}, one defines the edges as
\begin{eqnarray*}
00 &\to& 0_e; \quad 01 \to 1_e \quad 12 \to 2_e; \quad 13 \to 3_e; \\
20 &\to& 4_e; \quad 21 \to 5_e \quad 32 \to 6_e; \quad 33 \to 7_e\;.
\end{eqnarray*}
A suitable set of independent edge staying rates is 
$q_{0_e},q_{2_e},q_{5_e},q_{7_e}$ and I will drop the subscript $e$ as long
as there is no confusion with the vertex code. The remaining edge and
vertex staying rates for periodic orbits of period $n$ can be written in 
terms of the edge staying rates above with the help of
Eqs.\ (\ref{cons-q}), i.e., one obtains
\begin{eqnarray} \label{esr-4}
q_1 = q_4 &=& \frac{1}{4} \left( n - q_0 + q_2 - 3 q_5 - q_7\right)\\ \nonumber
q_3 = q_6 &=& \frac{1}{4} \left( n - q_0 - 3 q_2 + q_5 - q_7\right)
\end{eqnarray} 
for the edge rates and 
\begin{eqnarray} \label{vsr-4}
\tilde{q}_0&=&\frac{1}{4} \left(n+3 q_0+q_2-3 q_5-q_7\right)\\ \nonumber
\tilde{q}_1= \tilde{q}_2
&=&\frac{1}{4} \left(n - q_0+q_2+q_5-q_7\right)\\ \nonumber
\tilde{q}_3&=&\frac{1}{4} \left(n-q_0-3 q_2 + q_5 + 3 q_7\right)
\end{eqnarray} 
for the vertex staying rates $\tilde{q}_i$ and the index $i$ denotes the 
vertex code, here.\\

The periodic orbit length degeneracy function 
$P_n(q_{0},q_{2},q_{5},q_{7})$ can be computed by starting from 
\[ P_n(0,\frac{n}{2},\frac{n}{2},0) = 2 \qquad \mbox{for} 
\quad n \;\; \mbox{even}\; ;\]
the set of edge staying rates above corresponds to the orbits 
of length $n$ with edge symbol code
\be{po-4} 2\; 5\; 2\; 5\ldots 2\; 5 \quad \mbox{and} \quad  5\; 2\; 5\ldots 2\; 5\; 2 \; .\ee
One proceeds by noting that an edge symbol `2' in
the sequences (\ref{po-4}) can be replaced by the sequence `3 6' to
give an orbit of length $n+1$. Similarly one may substitute a symbol
'5' by the sequence `4 1'. Replacing $m$ symbols '2' and $k$ symbols 
'5', $m,k \le \frac{n}{2}$, one obtains
\[
P_{n+m+k}(0,\frac{n}{2} -m, \frac{n}{2} - k, 0) = 
2 \bino{\frac{n}{2}}{m} \bino{\frac{n}{2}}{k} 
+ \bino{\frac{n}{2}-1}{m-1} \bino{\frac{n}{2}}{k} 
+ \bino{\frac{n}{2}}{m} \bino{\frac{n}{2}-1}{k-1}, 
\]
and the last two terms in the sum come from orbits which start with
a symbol `6' or a symbol `1', respectively. After replacing 
$n + m +k $ by the new periodic orbit length $n'$, i.e., 
$n = n' - m - k$ and writing 
$\tilde{q}_1 = \frac{1}{4} (n' + q_2 + q_5)$ with 
$q_2 = \frac{1}{2}(n'-3 m -k)$, $q_5 = \frac{1}{2}(n'-m - 3 k)$ 
one obtains
\be{deg-4-00}
P_{n'}(0,q_2, q_5, 0) = 
2 \bino{\tilde{q}_1}{m} \bino{\tilde{q}_1}{k} + 
  \bino{\tilde{q}_1-1}{m-1} \bino{\tilde{q}_1}{k} + 
  \bino{\tilde{q}_1}{m} \bino{\tilde{q}_1-1}{k-1}\, .
\ee
Next, one notes that an edge symbol `0' or `7' can be inserted between
any symbol `4' and `1' or `3' and `6', respectively, to obtain a periodic
orbit of length $n'+1$. Inserting $q_0$ symbols `0' and $q_7$ 
symbols `7' into $k$ sequences '4 1' and $m$ sequences `3 6' 
(with repetition) leads to 
\begin{eqnarray*}
P_{n'+q_0+q_7}(q_0,q_2, q_5, q_7) &=&
2 \bino{\tilde{q}_1}{m} \bino{\tilde{q}_1}{k} 
\bino{m+q_7 -1}{q_7} \bino{k+q_0 -1}{q_0}\\
&+&  \bino{\tilde{q}_1-1}{m-1} \bino{\tilde{q}_1}{k} 
\bino{m+q_7}{q_7} \bino{k+q_0 -1}{q_0}\\
&+&\bino{\tilde{q}_1}{m} \bino{\tilde{q}_1-1}{k-1}
\bino{m+q_7-1}{q_7} \bino{k+q_0}{q_0}\, .
\end{eqnarray*}
After rescaling to the new periodic orbit length $n'' = n' + q_0 + q_7$ 
and summing the three contributions, one obtains
\be{deg-4-qq}
P_{n''}(q_0,q_2, q_5, q_7) = \frac{n''}{\tilde{q}_1}
\bino{\tilde{q}_1}{m} \bino{\tilde{q}_1}{k} 
\bino{m+q_7 -1}{q_0} \bino{k+q_0 -1}{q_7}
\ee
with $\tilde{q}_1 = \frac{1}{4}(n'' - q_0 +q_2 + q_5 - q_7)$ as in 
(\ref{vsr-4}). The final result (\ref{def-fun-4}) is obtained after 
noting that $m = q_3 = q_6 = \tilde{q}_1 - q_2$ and $k = q_1 = q_4 =
\tilde{q}_1 - q_5$. Special care has to be taken in the case 
$q_0$ or $q_7 = 0$.\\

\noindent
\underline{The case $N=6$}:\\[.3cm]
The periodic orbit length degeneracy function for $N =6$ can be derived 
by ideas similar to those outlined for $N = 4$; I will sketch the main
steps here and leave the details to the reader.

An edge symbol code is defined starting from the vertex
symbol code used in Fig.\ \ref{Fig:graph6} to be
\begin{eqnarray*}
0\,0 &\to& 0_e; \quad 0\,1 \to 1_e \quad 1\,2 \to 2_e; \quad 1\,3 \to 3_e; \\
2\,4 &\to& 4_e; \quad 2\,5 \to 5_e \quad 3\,0 \to 6_e; \quad 3\,1 \to 7_e ;\\
4\,2 &\to& 8_e; \quad 4\,3 \to 9_e \quad 5\,4 \to 10_e; \quad 5\,5 \to 11_e\;.
\end{eqnarray*}
A suitable set of independent edge staying rates is $q_{0_e},q_{3_e},
q_{4_e},q_{7_e}, q_{8_e}, q_{11_e}$ and I will drop the subscript $e$ from
now on. The other edge staying rates of orbits of period $n$ are then given 
by
\begin{eqnarray} \label{esr-6}
q_1 = q_6 &=& \frac{1}{6} 
\left( n - q_0 + 3 q_3 + q_4 - 5 q_7 -3 q_8 - q_{11}\right)\\ \nonumber
q_2 &=& \frac{1}{6} 
\left( n - q_0 - 3 q_3 + q_4 + q_7 -3 q_8 - q_{11}\right)\\ \nonumber
q_5 = q_{10} &=& \frac{1}{6} 
\left( n - q_0 - 3 q_3 - 5 q_4 + q_7 +3 q_8 - q_{11}\right)\; ;
\end{eqnarray} 
the vertex rates are 
\be{vsr-6}
\tilde{q}_0 = q_0 + q_1; \quad \tilde{q}_1 = \tilde{q}_3 = q_2 + q_3;
\quad \tilde{q}_2 = \tilde{q}_4 = q_4 + q_5; \quad
\quad \tilde{q}_5 = q_{10} + q_{11}\; .
\ee

A suitable starting point for the periodic orbit length degeneracy 
function is the periodic orbit `2 4 9 7' (in edge code) or 
`1 2 4 3' in vertex code, see Fig.\  \ref{Fig:graph6}. One obtains
\[ P_n(0,0,q_4=\frac{n}{4},q_7= \frac{n}{4},0,0) = 4 \qquad \mbox{for} 
\quad n \,\mbox{mod}\, 4 = 0\; .\]
A symbol `7' can be followed by a loop `3 7' (with repetition), a
symbol `4' may be followed by a loop `8 4' (with repetitions). Inserting
$k$ loops `3 7' and $m$ loops '8 4' into a sequence `2 4 9 7 $\ldots$'
of length $n - 2k - 2m$ yields 
\begin{eqnarray*}
&P_n&(0,k,\frac{1}{4}(n - 2k + 2m),\frac{1}{4}(n + 2k - 2m),l,0) =\\
&2& \bino{\frac{1}{4}(n + 2k - 2m) -1}{k} 
\bino{\frac{1}{4}(n - 2k - 2m) -1}{m}
+ \bino{\frac{1}{4}(n + 2k - 2m)}{k} 
\bino{\frac{1}{4}(n - 2k - 2m) -1}{m}\\
&+& \bino{\frac{1}{4}(n + 2k - 2m) -1}{k} 
\bino{\frac{1}{4}(n - 2k - 2m)}{m}
+ \bino{\frac{1}{4}(n + 2k - 2m) -1}{k-1} 
\bino{\frac{1}{4}(n - 2k - 2m)-1}{m}\\
&+& \bino{\frac{1}{4}(n + 2k - 2m) -1}{k} 
\bino{\frac{1}{4}(n - 2k - 2m) -1}{m-1}\; ,
\end{eqnarray*}
and the different terms in the sum correspond to a first symbol
in the periodic orbit code being either `2' or `9', `7', `4', '3' or
`8', respectively. Next, one notes that every symbol '7` or
'4' can be replaced by the sequence '6 1' or '5 10', respectively.
I omit the somewhat lengthy combinatorial expressions here.
The full periodic orbit length degeneracy function is finally 
obtained after inserting symbols '0' or '11' into the sequences
'6 1' or `5 10', respectively, and summing over the various 
binomial coefficients.
\end{appendix}

\end{document}